\documentclass[preprint,aps,12pt,showpacs,nofootinbib,tightenlines]{revtex4}
\usepackage{amsmath}
\usepackage{amssymb}
\usepackage{CJK}
\usepackage{epsfig}
\usepackage{graphicx}
\usepackage{subfigure}
\usepackage{slashed}
\usepackage{multirow}
\newcommand{\met}{\not\!\!\!E_{T}}
\newcommand{\mttwo}{\ensuremath{m_\mathrm{T2}}}

\newcommand{\pTvec}{\mathbf{p}_\mathrm{T}}
\newcommand{\qTvec}{\mathbf{q}_\mathrm{T}}
\newcommand{\pT}{p_\mathrm{T}}
\newcommand{\qT}{q_\mathrm{T}}
\newcommand{\mT}{m_\mathrm{T}}

\begin{document}
\begin{CJK*}{GBK}{song}
\def\pslash{\rlap{\hspace{0.02cm}/}{p}}
\def\eslash{\rlap{\hspace{0.02cm}/}{e}}
\title {T-odd top partner pair production in the dilepton final states at the LHC in the littlest Higgs Model with
T-parity}
\author{Bingfang Yang$^{1}$}\email{yangbingfang@htu.edu.cn}
\author{Huaying Zhang$^{1}$}
\author{Biaofeng Hou$^{1}$}
\author{Ning Liu$^{2}$}
\affiliation{$^1$College of Physics and Materials Science, Henan Normal
	University, Xinxiang 453007, China\\
	$^2$Department of Physics and Institute of Theoretical Physics,
	Nanjing Normal University, Nanjing 210023, China
	\vspace*{1.5cm}  }

\begin{abstract}

In the littlest Higgs Model with T-parity, we discuss the pair
production of the T-odd top partner ($T_{-}$) which decays almost 100\% into the top
quark and the lightest T-odd particle ($A_{H}$). Considering the current constraints, we investigate the
observability of the T-odd top partner pair production through the process $pp\rightarrow T_{-}\bar{T}_{-}\rightarrow t\bar{t}A_{H}A_{H}$ in final states with two leptons at 14 TeV LHC. We analyze the signal
significance and find that the lower limit on the T-odd top partner mass are about 1.04 TeV, 1.14 TeV, 1.23 TeV at $2\sigma$ confidence level at 14TeV LHC with the integrated luminosity of 30 fb$^{-1}$, 100fb$^{-1}$, 300fb$^{-1}$. This lower limit can be raised to about 1.34(1.44) TeV if we use 1000(3000) fb$^{-1}$ of integrated luminosity.

\end{abstract}
 \maketitle
\section{Introduction}
\noindent The discovery of Higgs boson by the ATLAS\cite{higgsatlas} and
CMS\cite{higgscms} collaborations is a major milestone for theoretical and experimental particle physics. The
Standard Model (SM) is already hugely successful, but there are still some unresolved problems, one of them is the naturalness \cite{Naturalness}. As the two most important particles, the top quark and the Higgs boson play a key role therein\cite{top-higgs}. Around this problem, various theories beyond the SM have been proposed in the past decades. Among these theories, the littlest Higgs Model with
T-parity (LHT) \cite{LHT} is one of the popular candidates.

The little Higgs models construct the Higgs boson as a
pseudo-Nambu-Goldstone particle arising from a global symmetry at high scale and the LHT model is an attractive representative of these models. In the LHT model, the T-parity
partners cancel the one-loop quadratic divergence contributions to Higgs mass from the corresponding SM particles. Among these partners, the top partner is very intriguing since it is responsible for canceling the largest
quadratic divergence of the Higgs mass induced by the
top quark loop. For this reason, the relevant researches have been extensively
carried out\cite{LHT-tpartner}.

Recently, many searches for the vector-like top partner through the
pair or single production have been
performed at LHC\cite{LHC13}. 
The search for direct top squark pair production at 13 TeV
LHC have been performed by the ATLAS and CMS collaborations in
various final states, where the search in dilepton final states has
used 36.1 fb$^{-1}$ of integrated luminosity collected by the ATLAS
detector\cite{LHC13-2} and observes no evidence for an excess above
the expected background from SM processes. For
neutralino masses below 150 GeV, masses of the lightest top squark
below 700 GeV are excluded at 95\% confidence level. The
similar search has also been performed by the CMS collaboration with
12.9 fb$^{-1}$ data\cite{LHC13-3}. In other final states, the
exclusion limits of top squark masses are pushed higher.  Apart from direct searches,
the indirect searches for the top partners have been extensively
investigated\cite{indirectbound}. The null results of the top partners, in conjunction with the electroweak precision observables (EWPOs) and the recent Higgs data, have tightly constrained the
parameter space of the LHT model \cite{lht-fit}. 

The LHT model also predicts exotic top partner, which is T-odd under T-parity and need to be pair-produced. Note that this T-odd top partner does not decay to the standard patterns $Wb,ht$ and $Zt$, but will decay into the lightest odd state and SM particles, which will share the same signature with the stop quark pair production in the R-parity conserving supersymmetry. At the LHC, the relevant search has been performed independently in pair-produced exotic top
partners, each decay to an on-shell top (or antitop) quark and a
long-lived undetected neutral particle\cite{tm-LHC}. On the other hand, the relevant theoretical studies have also been done\cite{T-oddtop}. Especially, the signals of the T-odd
top partner pair production in fully hadronic channel \cite{hadronic} and semileptonic channel\cite{semileptonic} at the LHC have been studied before the discovery of the Higgs boson. In this work, we will focus on the search for the T-odd
top partner pair production in dilepton final states at the LHC.

The paper is organized as follows. In Sec.II we review the top
partner in the LHT model. In Sec.III we give the cross section of the T-odd top partner pair production at the 14TeV LHC under the current indirect constraints. In Sec.IV we investigate the signal and discovery potentiality of the T-odd top
partner pair production in the dilepton final states at the LHC. Finally, we draw our conclusions
in Sec.V.

\section{Top partner in the LHT model}\label{section2}

The LHT model is a non-linear $\sigma$ model based on the coset
space $SU(5)/SO(5)$. The global group $SU(5)$ is spontaneously
broken into $SO(5)$ at scale $f\sim \mathcal O$(TeV) by the vacuum
expectation value (VEV) of the $\Sigma$ field, $\Sigma_0$, which is
given by
\begin{eqnarray}
\Sigma_0=
\begin{pmatrix}
{\bf 0}_{2\times2} & 0 & {\bf 1}_{2\times2} \\
0 & 1 &0 \\
{\bf 1}_{2\times2} & 0 & {\bf 0}_{2\times 2}
\end{pmatrix}.
\end{eqnarray}
Concurrently, the VEV $\Sigma_0$ breaks the gauged subgroup $\left[ SU(2)
\times U(1) \right]^{2} $ of $SU(5)$ down to the diagonal SM
electroweak group $SU(2)_L \times U(1)_Y$. After the symmetry
breaking, there arise 4 new heavy gauge bosons
$W_{H}^{\pm},Z_{H},A_{H}$ whose masses given at $\mathcal
O(v^{2}/f^{2})$ by
\begin {equation}
M_{W_{H}}=M_{Z_{H}}=gf(1-\frac{v^{2}}{8f^{2}}),~~M_{A_{H}}=\frac{g'f}{\sqrt{5}}
(1-\frac{5v^{2}}{8f^{2}})
\end {equation}
with $g$ and $g'$ being the SM $SU(2)_L$ and $U(1)_Y$ gauge
couplings, respectively. The lightest
$T$-odd particle $A_{H}$ can serve as a candidate for dark matter (DM). In
order to match the SM prediction for the gauge boson masses, the VEV
$v$ needs to be redefined as
\begin{equation}
v = \frac{f}{\sqrt{2}} \arccos{\left( 1 -
\frac{v_\textrm{SM}^2}{f^2} \right)} \simeq v_\textrm{SM} \left( 1 +
\frac{1}{12} \frac{v_\textrm{SM}^2}{f^2} \right) ,
\end{equation}
where $v_\textrm{SM}$ = 246 GeV.

For each SM quark, the implementation of T-parity requires
the existence of mirror partners with T-odd quantum number. 
In the top quark sector, an additional T-even heavy
top partner $T_{+}$ is introduced to cancel the large one-loop quadratic
divergences caused by the top quark in order to stabilize the Higgs mass. Meanwhile, the implementation
of T-parity requires also a T-odd mirror partner $T_{-}$.
The top partner $T_{+}$ mixes with the SM top quark and leads to a
correction of the top quark couplings with
respect to the SM values. The
mixing can be parameterized by dimensionless ratio
$R=\lambda_1/\lambda_2$, where $\lambda_1$ and $\lambda_2$ are two
dimensionless top quark Yukawa couplings. 
The Yukawa term generates the masses of the top quark and its
partners, which are given at $\mathcal O(v^{2}/f^{2})$ by
\begin{eqnarray}
&&m_t=\frac{\lambda_2 v R}{\sqrt{1+R^2}} \left[ 1 + \frac{v^2}{f^2}
\left( -\frac{1}{3} + \frac{1}{2} \frac{R^2}{(1+R^2)^2} \right)\right]\nonumber \\
&&m_{T_{+}}=\frac{f}{v}\frac{m_{t}(1+R^2)}{R}\left[1+\frac{v^{2}}{f^{2}}\left(\frac{1}{3}-\frac{R^2}{(1+R^2)^2}\right)\right] \nonumber \\
&&m_{T_{-}}=\frac{f}{v}\frac{m_{t}\sqrt{1+R^2}}{R}\left[1+\frac{v^{2}}{f^{2}}\left(\frac{1}{3}-\frac{1}{2}\frac{R^2}{(1+R^2)^2}\right)\right]\label{Tmass}
\end{eqnarray}
Since the $T_{+}$ mass is always larger than the
$T_{-}$ mass, the $T_{+}$ can decay into $A_{H}T_{-}$ in addition to
the conventional decay modes ($Wb, ht, Zt$). The T-odd top partner
$T_{-}$ has a simple decay pattern, which decays almost 100\% into
the $A_{H}t$ mode.

\section{T-odd top partner pair production at the LHC}
 \noindent 

\begin{figure}[htbp]
\begin{center}
\scalebox{0.35}{\epsfig{file=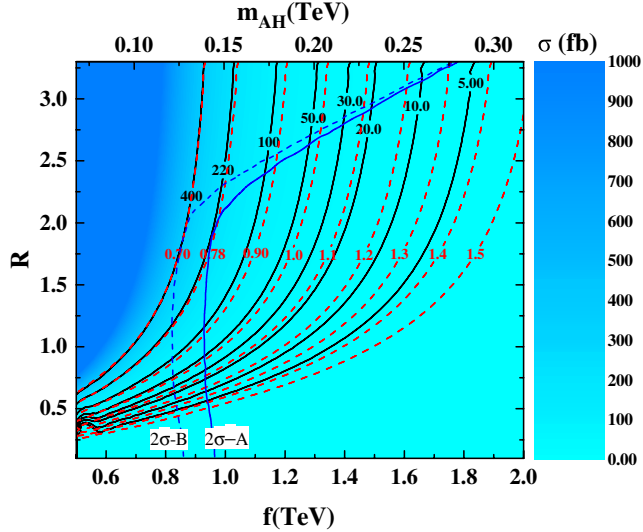}}\vspace{-0.5cm}\hspace{-0.cm}
\caption{The production cross section of $pp\to T_{-}\bar{T}_{-}$ in the $R\sim f$ plane at the 14TeV LHC, where the black solid lines correspond to the typical cross sections, the magenta solid(dash) lines correspond to 2$\sigma$ exclusion limits for Case A(B), the red dot lines correspond to the T-odd top partner masses (in units of TeV). }\label{mass}
\end{center}
\end{figure}

In Fig.\ref{mass}, we show the production cross section of process $pp\to T_{-}\bar{T}_{-}$ in the $R\sim f$ plane at the 14TeV LHC. For clarity,  we also show the typical T-odd top partner masses and the 2$\sigma$ exclusion limits from the indirect measurements in this plane. Here, the indirect constraints on the T-odd top partner mass including the latest Higgs data, EWPOs and
$R_{b}$ in our previous
work\cite{prework} have been updated by the package
\textsf{HiggsSignals-2.1.0}\cite{HiggsSignals}and
\textsf{HiggsBounds-5.1.0}\cite{HiggsBounds}. We can see that the
combined constraints can respectively exclude the scale $f$ up to $930(820)\rm GeV$ and $m_{T_{-}}$ up to
$780(700)\rm GeV$ at $2\sigma$ confidence level for Case A(B).

Here, the Case A and Case B denote two
possible ways to construct the T-invariant Yukawa interactions of the
down-type quarks and charged leptons\cite{caseAB}. In the two cases, the corrections to the
Higgs couplings with
respect to their SM values are given at order $\mathcal{O} \left(
v_{SM}^4/f^4 \right)$ by ($d \equiv d,s,b,\ell^{\pm}_i$)
\begin{eqnarray}
\frac{g_{h \bar{d} d}}{g_{h \bar{d} d}^{SM}} &=& 1-
\frac{1}{4} \frac{v_{SM}^{2}}{f^{2}} + \frac{7}{32}
\frac{v_{SM}^{4}}{f^{4}} \qquad \text{Case A} \nonumber \\
\frac{g_{h \bar{d} d}}{g_{h \bar{d} d}^{SM}} &=& 1-
\frac{5}{4} \frac{v_{SM}^{2}}{f^{2}} - \frac{17}{32}
\frac{v_{SM}^{4}}{f^{4}} \qquad \text{Case B}
\label{dcoupling}
\end{eqnarray}

These two cases do not differ in the collider phenomenology of the LHT model and only arise differences in the discussion of constraints from the Higgs sector and EWPO as shown in Fig.\ref{mass}. So, we will focus on the Case A in the study of T-odd top partner pair production. Besides, the heavy photon $A_{H}$ is the DM candidate, which will be constrained by the relic density. According to our previous work\cite{DMLHT}, $A_{H}$ needs to co-annihilate with the T-odd mirror fermions and the masses of both need to be approximatively degenerate to give the correct DM relic density. One should note that the measured DM relic density has no impact on the phenomenology of this work.

Recently, the latest research with the LHC-13 TeV data has been performed and found that the scale $f$ below 950 GeV for Case A can be excluded at $2\sigma$ confidence level\cite{LHT13TeV}. We can see that the bound on the scale $f$ with the available 13 TeV results only improves little compared to the 8 TeV results.  Moreover, the bounds on the top partner masses are not given explicitly in Ref.\cite{LHT13TeV} due to the fixed selection of the parameter $R$, and this will be the focus of this paper.
Furthermore, we can see that the cross section of the process $pp\to T_{-}\bar{T}_{-}$ depends almost entirely on the T-odd top partner mass and decreases rapidly with the increase of this mass. Considering the 2$\sigma$ exclusion limits, the cross sections can maximally reach 220(400)fb for Case A (B).

\section{SIGNAL AND DISCOVERY POTENTIALITY}

\begin{figure}[htbp]
	\scalebox{0.55}{\epsfig{file=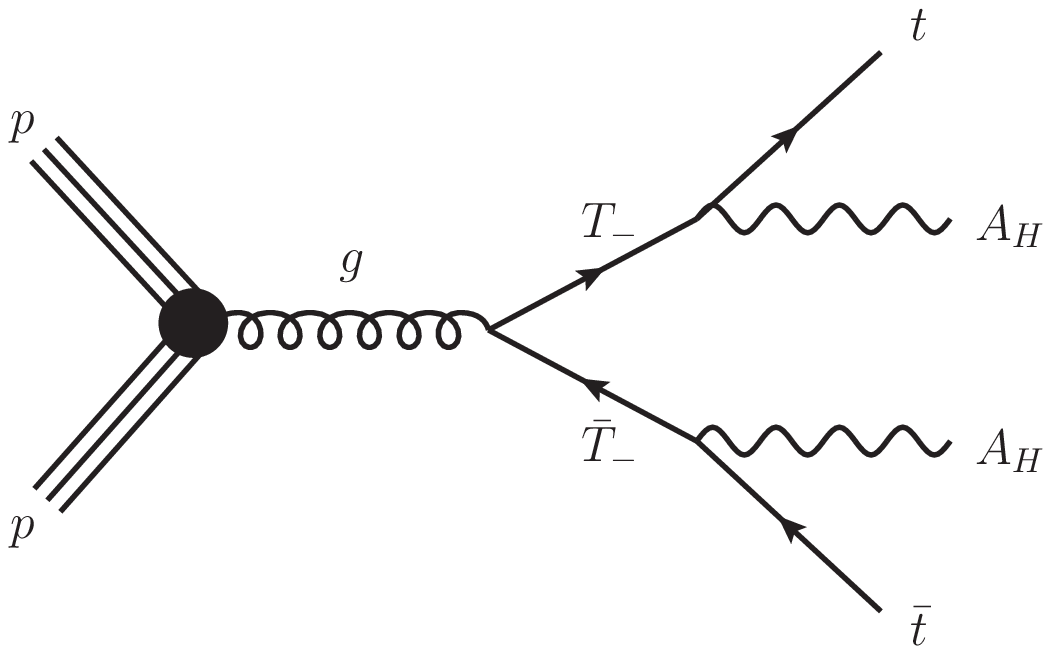}}\vspace{-0.5cm}\hspace{0.5cm}
	\scalebox{0.55}{\epsfig{file=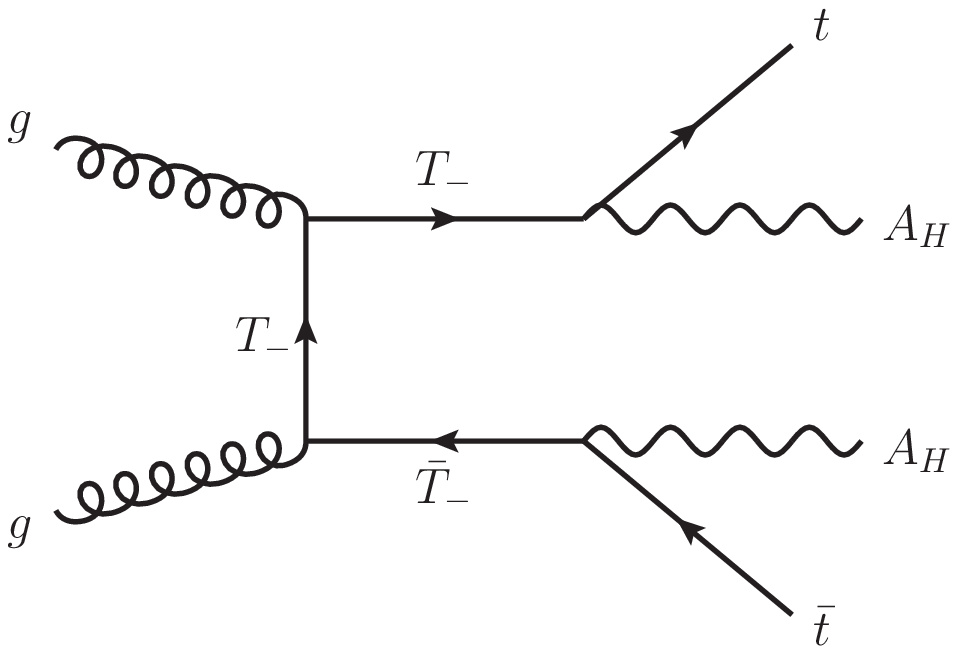}}
	\caption{Exemplary feynman diagrams of the production and decay of
		the T-odd vector-like top quark pair at the LHC in the LHT
		model.}\label{pptt}
\end{figure}

In Fig.\ref{pptt}, we show the exemplary feynman diagrams of the
production and decay of the T-odd vector-like top quark pair at the
LHC. We can see that the leading production mechanism for the T-odd top partner pair is via QCD interactions.

In the next section, we will perform the Monte Carlo simulation and
explore the sensitivity of the T-odd top partner pair production through
the channel,
\begin{eqnarray}
pp\rightarrow T_{-}\bar{T}_{-}\rightarrow t(\rightarrow
l^{+}\nu_{l}b)\bar{t}(\rightarrow
l^{-}\bar{\nu_{l}}\bar{b})A_{H}A_{H}\rightarrow l^{+}l^{-}+2b+\met
\end{eqnarray}
which implies that the events contain one pair of oppositely charged
leptons $l^{+}l^{-}(l=e,\mu)$ with high transverse momentum, two
high transverse momentum $b$-jets and large missing transverse
energy $\met$.

For this signal, the dominant background arises from $pp\rightarrow
t\bar{t}$ in the SM, and the most relevant backgrounds come
from $tW$, $t\bar{t}V$($V=W,Z$) and $VV$($WW,WZ,ZZ$). We generate
the signal and background events by \textsf{MadGraph
	5}\cite{MadGraph} and use CTEQ6L as the parton distribution
functions (PDF). When generating the parton level events, the
renormalization and factorization scales are
set dynamically by default.
The cross sections of $t\bar{t}$ and $tW$ production are normalized to their NNLO+NNLL values\cite{tt-tw}, and the cross sections of $t\bar{t}V$ and $VV$ production are normalized to their NLO values\cite{MadGraph}.

The basic cuts are chosen as follows:
\begin{eqnarray}\label{basic}
\nonumber\Delta R_{ij} &>&  0.4\ ,\quad  i,j =  \ell, b~\text{or}\ j  \\
\nonumber  p_{T}^\ell &>& 10 \ \text{GeV}, \quad  |\eta_\ell|<2.5  \\
\nonumber  p_{T}^{b} &>& 20 \ \text{GeV}, \quad  |\eta_{b}|<2.5  \\
\nonumber  p_{T}^j &>& 20 \ \text{GeV},\quad  |\eta_j|<5.
\end{eqnarray}

We fed the events into \textsf{PYTHIA}\cite{PYTHIA} for parton showering and hadronization, and performed a fast detector simulations with \textsf{Delphes}\cite{Delphes}. The $b$-jet tagging
efficiency is taken as default value in delphes, where it is
parameterized as a function of the transverse momentum and rapidity
of the jets. \textsf{FastJet}\cite{fastjet} is used to cluster the
jets by choosing the anti-$k_{t}$ algorithm \cite{algorithm} with
distance parameter $\Delta R =0.4$.

\begin{figure}[htbp]
\begin{center}
\scalebox{0.4}{\epsfig{file=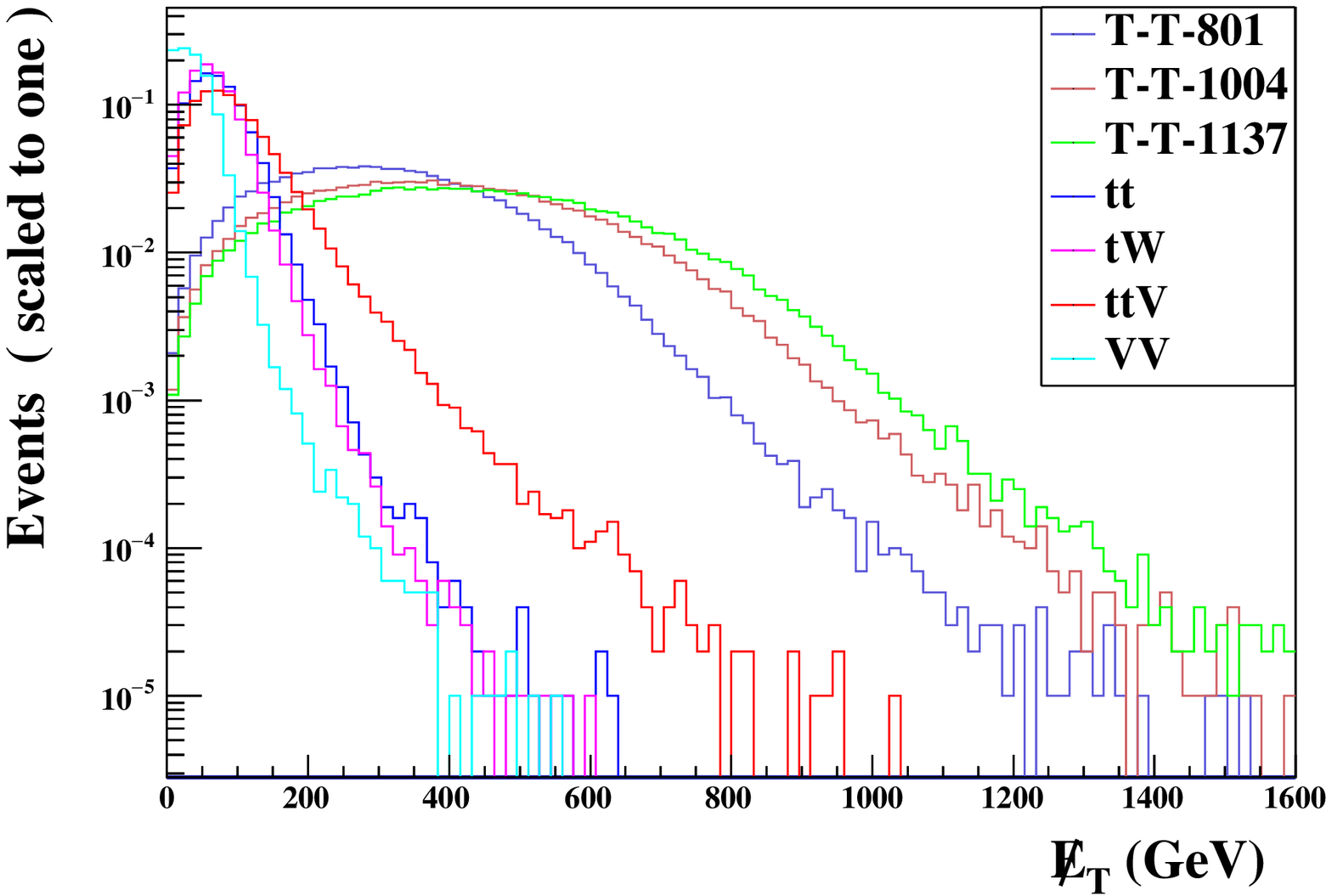}}\hspace{-0.3cm}
\scalebox{0.4}{\epsfig{file=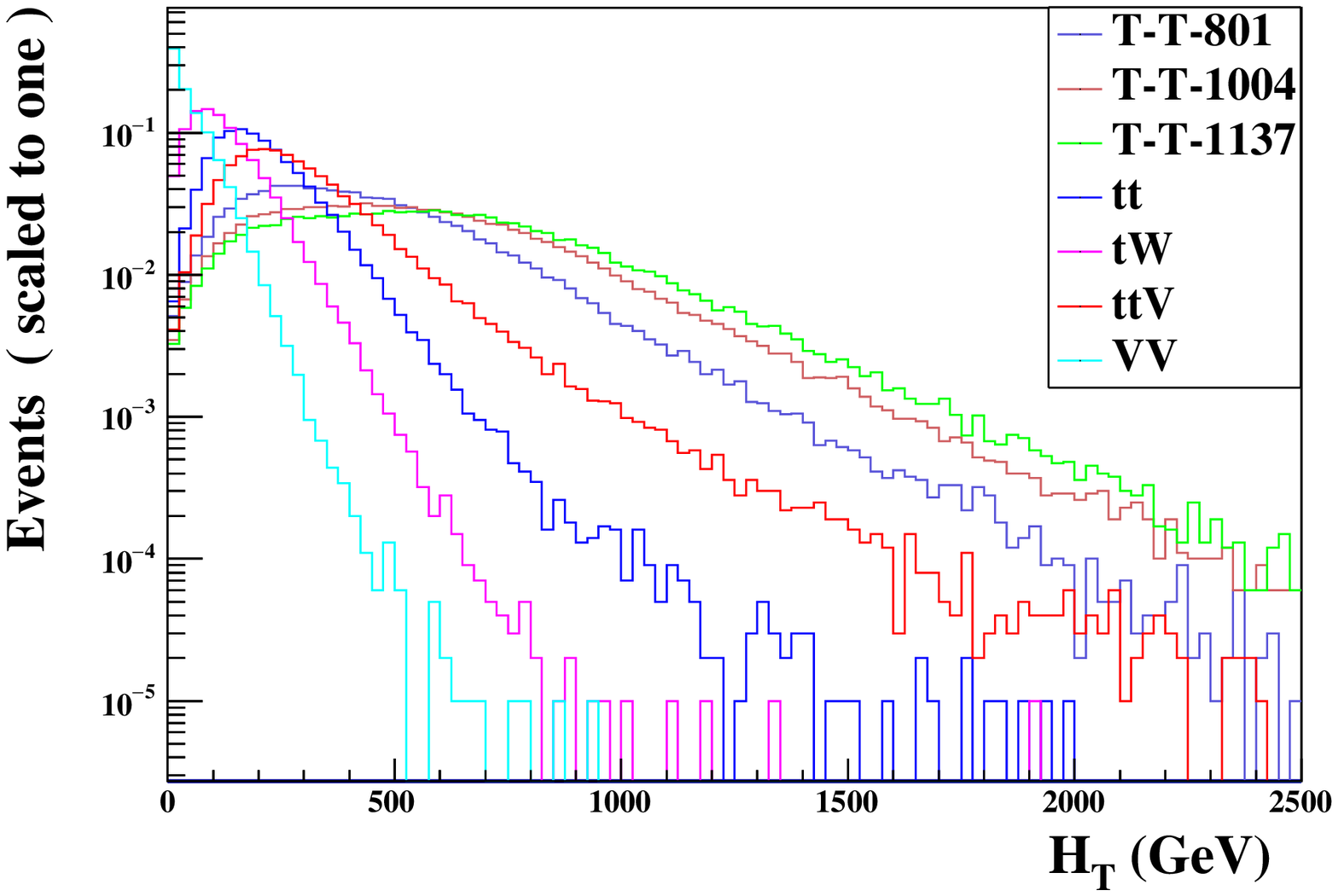}}\vspace{-0.5cm}
\caption{Normalized distributions of $\met$, $H_{T}$ in the signal
and backgrounds for the three signal benchmark points at 14 TeV LHC.
}\label{tmtm}
\end{center}
\end{figure}

The SM parameters are taken as follows\cite{parameter}
\begin{eqnarray}
\nonumber
\sin^{2}\theta_{W}=0.231,~\alpha_{e}=1/128,~M_{Z}=91.1876\textrm{GeV},~m_{t}=173.5\textrm{GeV},~m_{H}=125\textrm{GeV}.
\end{eqnarray}

Taking into account the uncertainty of measurements, we relax the
constraints on the T-odd top partner mass slightly and take $f=1000$ GeV, $R=2$ (correspond to $m_{T_{-}}=801$ GeV), $f=1000$ GeV, $R=1$ (correspond to $m_{T_{-}}=1004$ GeV), $f=1000$ GeV and $R=0.8$
(correspond to $m_{T_{-}}=1137$ GeV) for three benchmark points, and now the heavy photon mass is
$m_{A_{H}}=$150 GeV. In order to reduce the background and enhance
the signal contribution, some cuts of kinematic distributions are
needed. Since the dominant background arises from $t\bar{t}$, the
cuts should centered
around the $t\bar{t}$ to suppress the backgrounds. In Fig.\ref{tmtm}, we
show the normalized distributions of the missing transverse energy $\met$ and total transverse
energy $H_{T}$ in the
signal and backgrounds for the three signal benchmark points at 14 TeV LHC. Firstly, we can choose the large 
$\met$ cut to reduce the backgrounds. Then, the $H_{T}$ distribution can be also utilized to
remove the $t\bar{t}$ background obviously. 

\begin{figure}[htbp]
	\begin{center}
		\scalebox{0.35}{\epsfig{file=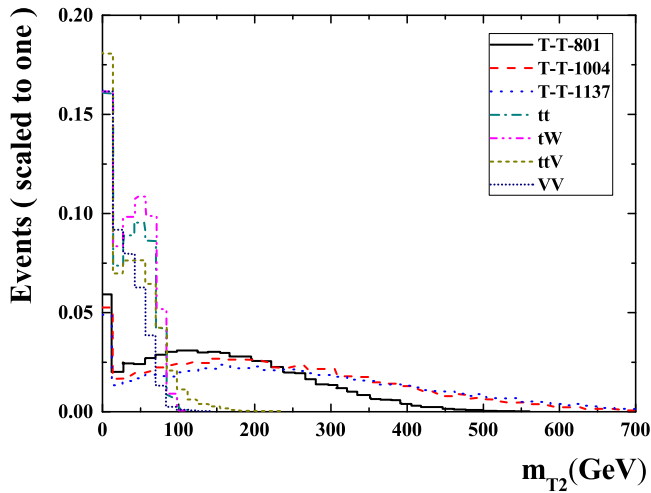}}\vspace{-0.5cm}
		\caption{Normalized distributions of $m_{T2}$ in the signal and
			backgrounds for the three signal benchmark points at 14 TeV LHC.
		}\label{mt2}
	\end{center}
\end{figure}

Besides, for this analysis the `stransverse' mass \mttwo~\cite{MT2}
is an effective kinematic variable, which has been suggested or used
in top-quark mass measurements and search for the supersymmetric
particles at the LHC. This quantity is defined as
\begin{equation*}
\mttwo ( \mathbf p^{\ell_1}_{\mathrm{T}}, \mathbf
p^{\ell_2}_{\mathrm{T}}, \mathbf p^{\mbox{{\footnotesize
miss}}}_{\mathrm{T}}) = \min_{\mathbf q_{\mathrm{T}} + \mathbf
r_{\mathrm{T}} = \mathbf p^{\mathrm{miss}}_{\mathrm{T}} } \left\{
\max [\;
    m_{\mathrm{T}}( \mathbf p^{\ell_1}_{\mathrm{T}}, \mathbf q_{\mathrm{T}} ),
    m_{\mathrm{T}}( \mathbf p^{\ell_2}_{\mathrm{T}}, \mathbf r_{\mathrm{T}}
    )
\;] \right\} ,
\end{equation*}
where $m_{\mathrm T}$ indicates the transverse mass, which is
defined by
\[
\mT(\pTvec,\qTvec) = \sqrt{2(\pT\qT-\pTvec\cdot\qTvec)}.
\]
$\mathbf p^{\ell_1}_{\mathrm{T}}$ and $\mathbf
p^{\ell_2}_{\mathrm{T}}$ are the transverse momenta of the two
leptons, and $\mathbf q_{\mathrm{T}}$ and $\mathbf r_{\mathrm{T}}$
are vectors which satisfy $\mathbf q_{\mathrm{T}} + \mathbf
r_{\mathrm{T}} = \mathbf p^{\mbox{\footnotesize
miss}}_{\mathrm{T}}$. The minimization is performed over all the
possible decompositions of $\mathbf p^{\mbox{\footnotesize
miss}}_{\mathrm{T}}$. We show the normalized distributions of
$m_{T2}$ in the signal and backgrounds for the three signal benchmark
points at 14 TeV LHC in Fig.\ref{mt2}.

We use \textsf{CheckMATE-1.2.2} \cite{CheckMATE} for analysis and apply charged lepton number $N(l)\geq 2$ to trigger the signal events after the basic cuts. According to the behavioral characteristics of above distributions,
events are selected to satisfy the following cuts:
\begin{eqnarray}
&&\textrm{Cut-1}:~N(l)\geq 2; \nonumber\\
&&\textrm{Cut-2}:~\met>160\textrm{GeV}; \nonumber\\
&&\textrm{Cut-3}: H_{T}>200\textrm{GeV};\nonumber\\
&&\textrm{Cut-4}: m_{T2}>120\textrm{GeV}.\nonumber
\end{eqnarray}
For clarity, we summarize the cut-flow cross sections of the signal and backgrounds at 14 TeV LHC in Table \ref{tab2}.

\begin{table}[ht]
\caption{Cut flow of the cross sections for the signal and the
backgrounds for the three signal benchmark points  $m_{T_{-}}=801, 1004, 1137$ GeV at 14 TeV LHC.
\label{tab2}}
\bigskip
\begin{tabular}{|c|c|c|c|c|c|c|c|c|}
\hline
     \multirow{2}{*}{Cuts}& \multicolumn{3}{c|}{Signal(S)(fb)} &\multicolumn{4}{c|}{Backgrounds(B)(fb)}\\
\cline{2-8}
    &$T_{-}\bar{T}_{-}$(801) &$T_{-}\bar{T}_{-}$(1004) & $T_{-}\bar{T}_{-}$(1137) & $t\bar{t}$&  $tW$ &$t\bar{t}V$&$VV$ \\
\hline
    Basic cuts&6.72&1.58&0.702&39874&3486&30.03&86475 \\
\hline Cut-1&3.05&0.618&0.243&22226&2119&16.82&3912\\    
\hline Cut-2&2.44&0.53&0.22&677.9&41.5&2.31&20.58\\
\hline
   Cut-3&1.58&0.364&0.15&393&13.8&1.62&0.18\\
\hline
   Cut-4&1.10&0.27&0.12&0.0&0.03&0.22&0.18 \\
\hline
\end{tabular}
\end{table}

\begin{figure}[htbp]
	\begin{center}
		\scalebox{0.3}{\epsfig{file=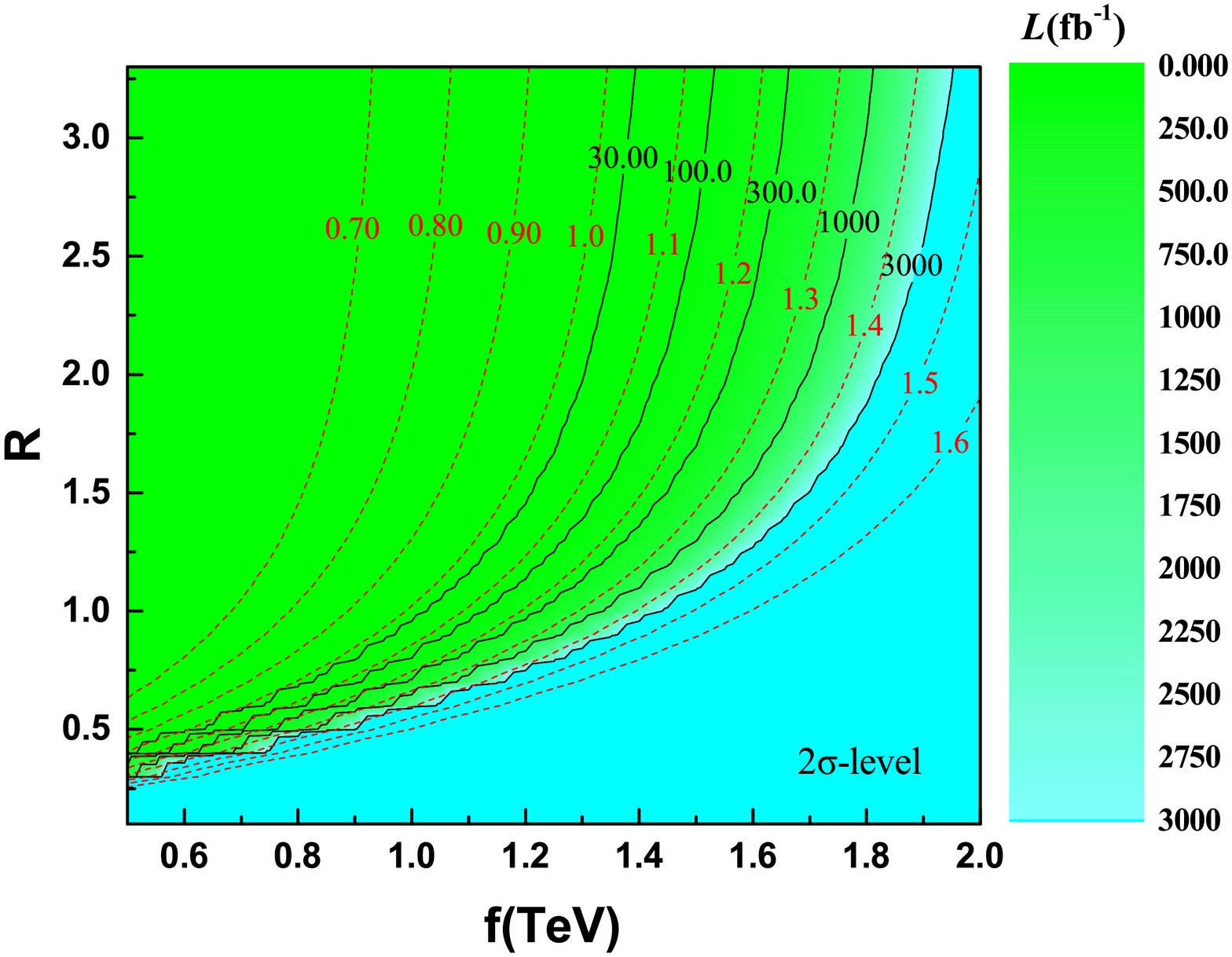}}\vspace{-0cm}\hspace{-0.cm}
				\scalebox{0.3}{\epsfig{file=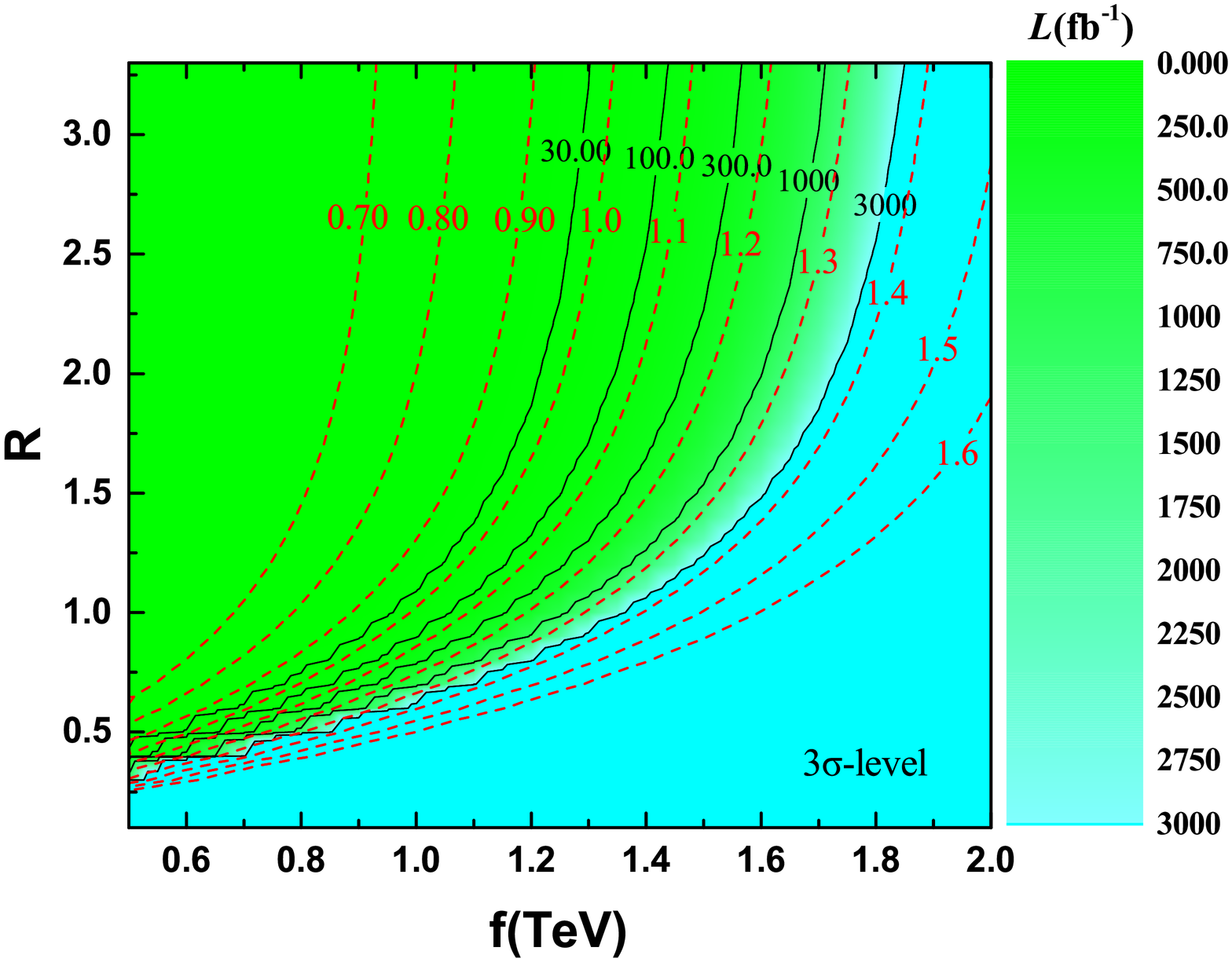}}\vspace{-0cm}\hspace{-0.cm}
		\caption{Excluded regions at $2\sigma$ and $3\sigma$ level depending on integrated luminosity in the $R\sim f$ plane at 14TeV LHC, where the red dot lines correspond to the $m_{T_{-}}$ (in units of TeV).}\label{ssl}
	\end{center}
\end{figure}

We can see that the total cut efficiency of the signal can reach $16.4\%$, $17.1\%$, $17.1\%$ for $m_{T_{-}}=801,1004,1137$ GeV, respectively. To estimate the observability
quantitatively, the Statistical Significance ($SS$) is calculated
after final cut by using Poisson formula\cite{Poisson}
\begin{eqnarray}
SS=\sqrt{2L\left [ (S+B)\ln\left(1+\frac{S}{B}\right )-S\right ]},
\end{eqnarray}
where $S$ and $B$ are the signal and background cross sections and
$L$ is the integrated luminosity. We chose the conservative cut efficiency 16.5\% of the signal and the excluded regions at $2\sigma$ and $3\sigma$ level depending on integrated luminosity in the $R\sim f$ plane at 14TeV LHC are shown in Fig.\ref{ssl}. We can see that the $T_{-}$ mass can be excluded up to about 1.04 TeV, 1.14TeV, 1.23 TeV at $2\sigma$ level with the integrated luminosity of 30fb$^{-1}$, 100fb$^{-1}$, 300fb$^{-1}$, respectively. If the integrated luminosity can be raised to $L=1000(3000)$ fb$^{-1}$, the lower limit on the $T_{-}$ mass will be pushed up to about 1.34(1.44)TeV. 

\begin{figure}[htbp]
	\begin{center}
		\scalebox{0.3}{\epsfig{file=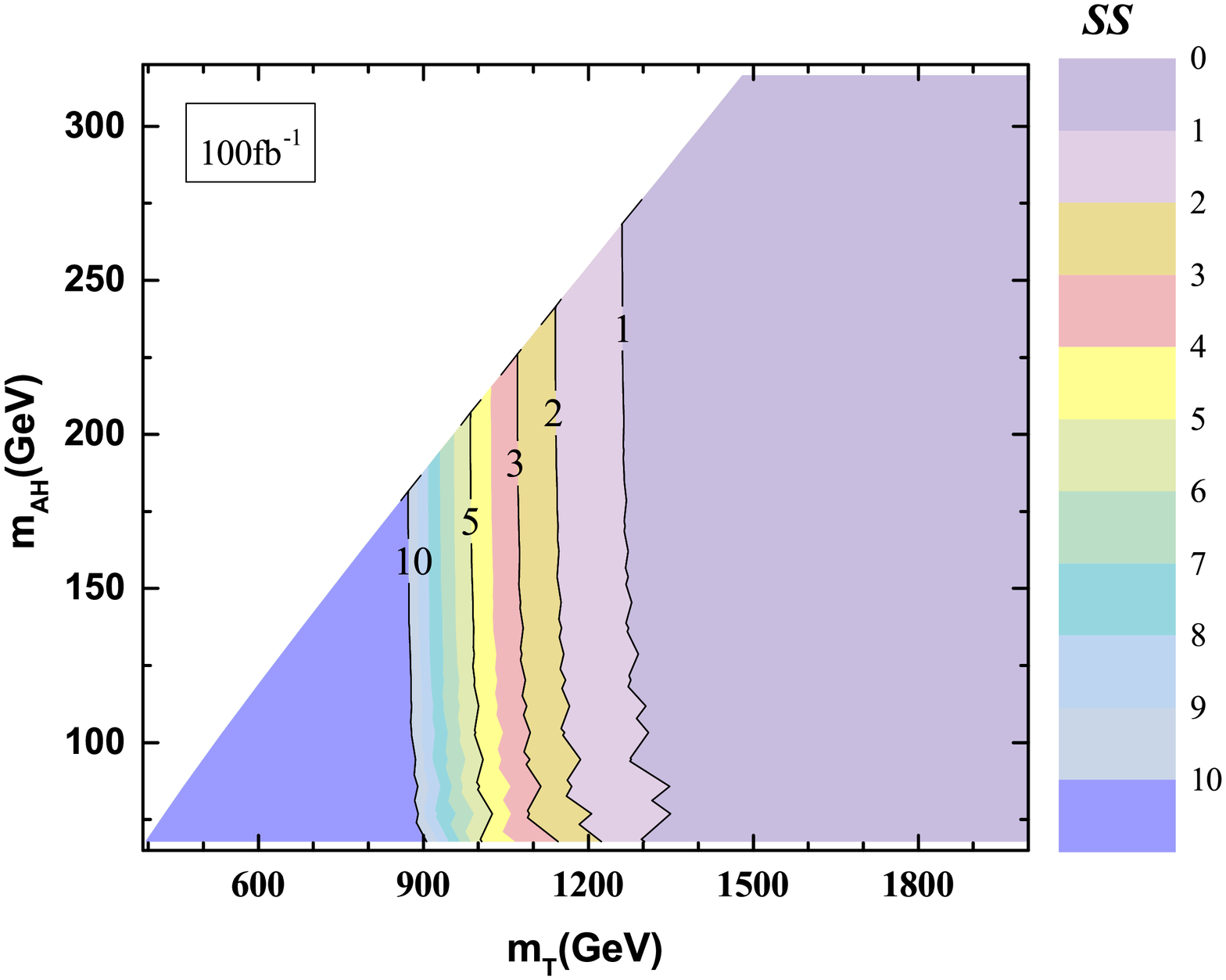}}\vspace{-0cm}\hspace{-0.cm}
		\scalebox{0.3}{\epsfig{file=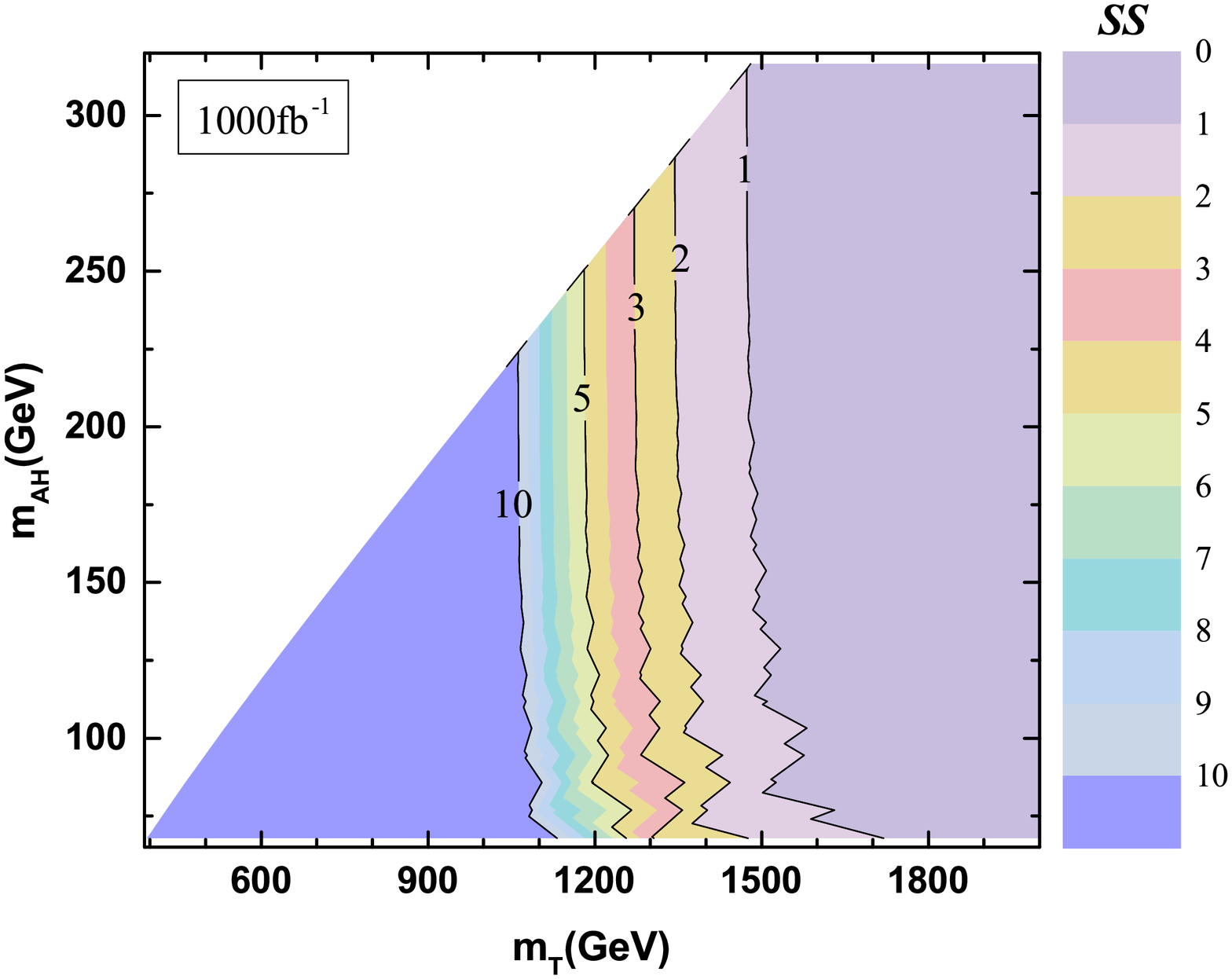}}\vspace{-0cm}\hspace{-0.cm}
		\caption{Contours of $SS$ with 100fb$^{-1}$ and 1000 fb$^{-1}$ of luminosity in the $m_{T}\sim m_{A_{H}}$ plane.}\label{ssl2}
	\end{center}
\end{figure}

In order to compare our results with that obtained in other modes, we also display the contours of $SS$ with 100fb$^{-1}$ and 1000 fb$^{-1}$ of luminosity in the $m_{T}\sim m_{A_{H}}$ plane in Fig.\ref{ssl2}. We can see that our limit on the $T_{-}$ mass is weaker than that in the fully hadronic mode\cite{hadronic} and the semileptonic mode\cite{semileptonic} due to the smaller signal events.

\section{Conclusions}

In this paper, we discuss the T-odd top partner pair production at the LHC in the LHT model. Under the current constraints, we investigate
the observability of the T-odd top partner pair production through the process
$pp\rightarrow T_{-}\bar{T}_{-}\rightarrow t\bar{t}A_{H}A_{H}$ with
two leptons in final states. We display the excluded regions at $2\sigma$ and $3\sigma$ level depending on integrated luminosity in the $R\sim f$ plane at 14 TeV LHC and find that the T-odd top partner mass $m_{T_{-}}$ can be excluded up to about 1.04 TeV, 1.14 TeV, 1.23 TeV at $2\sigma$ level with the integrated luminosity of 30fb$^{-1}$, 100fb$^{-1}$, 300fb$^{-1}$, respectively. This lower limit can be enhanced to about 1.34(1.44) TeV using 1000(3000) fb$^{-1}$ of integrated luminosity.

\section*{Acknowledgement}
We would like to thank Liangliang Shang for helpful discussions on
the CheckMATE package. This work is supported by the National
Natural Science Foundation of China (NNSFC) under grant Nos.
11405047, 11305049 and by the Startup Foundation
for Doctors of Henan Normal University under Grant No. qd15207.

\end{CJK*}

\vspace{0.5cm}

\begin{thebibliography}\\
\bibitem{higgsatlas}G. Aad et al. [ATLAS Collaboration], Phys. Lett. B 716, 1
(2012).
\bibitem{higgscms}S. Chatrchyan et al. [CMS Collaboration], Phys. Lett. B 716, 30
(2012).

\bibitem{Naturalness} L. Susskind, Phys.
Rev. D 20, 2619 (1979).

\bibitem{top-higgs}see examples: J.~Li, Z.~G.~Si, L.~Wu and J.~Yue,
Phys.\ Lett.\ B 779, 72 (2018); J.~Cao, L.~Wang, L.~Wu and J.~M.~Yang, Phys.\ Rev.\ D 84, 074001 (2011); A.~Kobakhidze, L.~Wu and J.~Yue, JHEP 1410, 100 (2014).

\bibitem{LHT}H. C. Cheng and I. Low, JHEP 0309, 051 (2003); JHEP 0408, 061
(2004); I. Low, JHEP 0410, 067 (2004).

\bibitem{LHT-tpartner} N. Liu, L. Wu, B. F. Yang and M. C. Zhang, Phys. Lett. B 753, 664-669 (2016);  H. Y. Wang and B. F. Yang, Adv. High Energy Phys. 2017, 5463128 (2017); B. F. Yang, B. F. Hou, H. Y. Zhang, Nucl. Phys. B  929, 207 (2018)

\bibitem{LHC13}G. Aad et al. [ATLAS Collaboration], ATLAS-CONF-2016-013;
ATLAS-CONF-2016-101; ATLAS-CONF-2016-102; ATLAS-CONF-2017-015; S.
Chatrchyan et al. [CMS Collaboration], CMS PAS B2G-15-008; CMS PAS
B2G-16-001; CMS PAS SUS-16-029, CMS PAS SUS-16-049; CMS PAS
SUS-16-028; CMS PAS SUS-16-051.

\bibitem{LHC13-2}G. Aad et al. [ATLAS Collaboration],
ATLAS-CONF-2017-034, CERN-EP-2017-150.

\bibitem{LHC13-3} S. Chatrchyan et al. [CMS Collaboration], CMS
PAS SUS-16-027.


\bibitem{indirectbound}see examples: C. C. Han, A. Kobakhidze, N. Liu, L. Wu and B. F. Yang, Nucl. Phys. B 890, 388
(2015); Lorenzo Basso, Jeremy Andrea, JHEP 1502, 032 (2015); J. Reuter, M. Tonini, JHEP 1501, 088 (2015); 
Y. B. Liu, Phys. Rev. D 95, 035013 (2017); Y. B. Liu, Y. Q. Li, Eur.Phys.J. C 77, 654 (2017);  M. Chala, Phys.Rev. D 96, 015028 (2017).

\bibitem{lht-fit}J. Reuter, M. Tonini and M. de Vries, JHEP 1402, 053 (2014); B. F. Yang, G. F. Mi and N. Liu, JHEP 1410, 47 (2014).

\bibitem{tm-LHC}G. Aad et al. [ATLAS Collaboration], Phys. Rev. Lett. 108, 041805
(2012); G. Aad et al. [ATLAS Collaboration], JHEP 1211, 094 (2012).


\bibitem{T-oddtop}H. C. Cheng, I. Low, and L. T. Wang, 
Phys. Rev. D 74, 055001 (2006); A. Anandakrishnan, J. H. Collins, M. Farina, E. Kuflik, M. Perelstein, Phys. Rev. D 93, 075009 (2016).

\bibitem{hadronic}P. Meade and M. Reece, Phys. Rev. D 74, 015010 (2006).

\bibitem{semileptonic}T. Han, R. Mahbubani, D. G. E. Walker, L. T. Wang, JHEP 0905, 117 (2009). 


\bibitem{prework} L. Wu, B. F. Yang and M. C. Zhang, JHEP 12, 152 (2016); B. F. Yang, J. Z. Han and N. Liu, Phys. Rev. D
95, 035010 (2017); B. F. Yang, Z. Y. Liu and N. Liu, Chin. Phys. C 41 (4), 043103 (2017).

\bibitem{HiggsSignals}P. Bechtle, S. Heinemeyer, O. St{\aa}l, T. Stefaniak, G. Weiglein, Eur.Phys.J. C 74, 2711 (2014);
O. St{\aa}l, T. Stefaniak, PoS EPS-HEP 2013, 314 (2013); P. Bechtle,
S. Heinemeyer, O. St{\aa}l, T. Stefaniak, G. Weiglein, JHEP 1411,
039 (2014).

\bibitem{HiggsBounds}P. Bechtle, O. Brein, S. Heinemeyer, G. Weiglein, K. E. Williams, Comput. Phys. Commun. 181, 138-167 (2010); Comput. Phys. Commun. 182, 2605-2631(2011); P. Bechtle, O. Brein, S.
Heinemeyer, O. St{\aa}l, T. Stefaniak, G. Weiglein, K. Williams, PoS
CHARGED 2012, 024 (2012); Eur.Phys.J. C 74, 2693 (2014); P. Bechtle,
S. Heinemeyer, O. St{\aa}l, T. Stefaniak, G. Weiglein, Eur.Phys.J. C
75, 421 (2015).

\bibitem{caseAB}C. R. Chen, K. Tobe and C.-P. Yuan, Phys. Lett. B 640, 263 (2006).

\bibitem{DMLHT}L. Wu, B. F. Yang and M. C. Zhang, JHEP 12, 152 (2016).

\bibitem{LHT13TeV}D. Dercks, G. Moortgat-Pick, J. Reuter, S. Y. Shim, arXiv:1801.06499.

\bibitem{MadGraph}J. Alwall et al., JHEP 07, 079 (2014).

\bibitem{tt-tw}M. Czakon, P. Fiedler and A. Mitov,
Phys. Rev. Lett. 110, 252004 (2013);
M. Czakon and A. Mitov, JHEP 01, 080 (2013); M. Czakon and A. Mitov, JHEP 12, 054 (2012); P. B\"{a}rnreuther, M. Czakon and A. Mitov, Phys. Rev. Lett. 109, 132001 (2012); M. Cacciari, M. Czakon, M. Mangano, A. Mitov and P. Nason, Phys. Lett. B 710, 612 (2012); M. Czakon and A. Mitov, Comput. Phys. Commun. 185, 2930 (2014); N. Kidonakis, Phys. Rev. D 82, 054018 (2010).

\bibitem{PYTHIA}T. Sjostrand, S. Mrenna and P. Z. Skands, JHEP 0605, 026 (2006).
\bibitem{Delphes}J. de Favereau, et al., JHEP 1402, 057 (2014).
\bibitem{fastjet}M. Cacciari, G.P. Salam and G. Soyez, Eur. Phys. J. C 72, 1896 (2012).

\bibitem{algorithm}M. Cacciari, G. P. Salam, and G. Soyez, JHEP 04, 063 (2008).
\bibitem{parameter}C. Patrignani et al., (Particle Data Group), Chin. Phys. C 40 (10), 100001 (2016).
\bibitem{MT2}C. G. Lester and D. J. Summers, Phys. Lett.
B 463, 99 (1999); A. Barr, C. Lester, and P. Stephens, J. Phys. G
29, 2343 (2003).
\bibitem{CheckMATE}M. Drees et al., Comput. Phys. Commun. 187, 227 (2014).
J. S. Kim, D. Schmeier, J. Tattersall and K. Rolbiecki,
Comput.Phys.Commun. 196, 535-562 (2015).

\bibitem{Poisson}G. Cowan, K. Cranmer, E. Gross, and O. Vitells, Eur. Phys. J. C 71,
1554 (2011).

\end{thebibliography}
\end{document}